\documentclass{iopart}
\usepackage{epsfig}

\newcommand{\pT}{p_\perp}
\newcommand{\kTaveSq}{\mbox{$\left\langle k_\perp^2\right\rangle$}}

\newcommand{\pTN}[1]{p_{\perp_{#1}}}

\newcommand{\Pythia}{\textsc{Pythia}}

\newcommand{\UNIT}[1]{\mbox{$\,{\rm #1}$}}

\newcommand{\GeV}{\UNIT{GeV}}
\newcommand{\GeVc}{\UNIT{GeV/c}}

\newcommand{\proz}{\UNIT{\%}}
\newcommand{\SqrtS}[1]{\mbox{$\sqrt s=#1\GeV$}}

\begin{document}

\title{Quenching of high $p_\perp$ hadrons by (pre--)hadronic FSI at RHIC}

\author{W Cassing\dag, K Gallmeister\dag \ and C
Greiner\ddag\footnote[3]{Talk presented by C. Greiner}} 

\address{\ddag\ Institut f\"ur Theoretische Physik, %
  Universit\"at Giessen, %
  Heinrich--Buff--Ring 16, %
  D--35392 Giessen, %
  Germany}
\address{\ddag\ Institut f\"ur Theoretische Physik, %
  Universit\"at Frankfurt, %
  Robert-Mayer-Str.~8--10, %
  D--60054 Frankfurt, %
  Germany}

\begin{abstract}
  Recently we have  conjectured that (in-)elastic collisions of
  (pre-)\-hadronic high momentum states with the bulk of hadrons
  in the late fireball
  might substantially account for the attenuation of the high transverse
  momentum hadrons at RHIC. The potential hadronic attenuation
  has therefore to be
  addressed in further detail before definite conclusions on possible
  QCD effects in a deconfined QGP phase can be drawn with respect to
  the materializing jets. From our transport studies we find that
  the interactions of fully formed hadrons are
  practically negligible in central Au+Au collisions at
  \SqrtS{200} for $\pT\geq6\GeVc$, but have some importance
  for the shape of the ratio $R_{AA}$ at lower $\pT$
  ($\leq6\GeVc$). However, a significant part of the large suppression seen
  experimentally is attributed to inelastic interactions of 'leading'
  jet-like pre-hadrons with the dense hadronic environment.
  In addition, we also show within this scenario
  first (preliminary)  results of near-side and
  far-side correlations of high $\pT$ particles from central Au+Au collisions
  at $\sqrt{s}$ = 200 GeV.
  It turns out that the near-side correlations are unaltered -- in
  accordance with experiment --
  whereas the far-side correlations are suppressed by $\sim$ 60\%. Since
  the experimental observation is a nearly complete disappearance
  of the far-side jet there should be some additional and earlier partonic
  interactions in the dense and possibly colored medium.
\end{abstract}


\section{Motivation and Concepts}

Within the standard picture of `jet quenching' in a quark gluon
plasma (QGP) the later materialization of the partonic jets are assumed
to take place outside the fireball without any further
interactions with the late stage dense hadronic environment.
Hence, measurements of jets seem to offer a direct access to probe the
early stage when the deconfined matter is very dense.
In fact, the PHENIX \cite{PHENIX1} and STAR \cite{STAR1} collaborations
have reported a large relative suppression of hadron spectra for
transverse momenta $\pT$ above $\sim3-4\GeVc$ which might point
towards the creation of a QGP, since this suppression is not
observed in d+Au interactions at the same bombarding energy per
nucleon \cite{E1,E2}.

\begin{figure}
\begin{center}
\includegraphics[width=7cm]{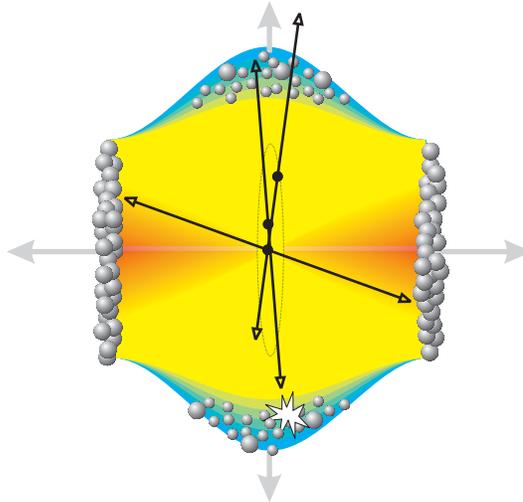}
\end{center}
\caption{Illustration of jets traveling through the
late hadronic stage of the reaction.
Only jets  from the region close to the initial surface can propagate
and fragment in the vacuum.
}
\label{fig:scheme}
\end{figure}

\begin{figure}
\begin{center}
\hspace*{\fill}
\includegraphics[angle=-90,width=6cm]{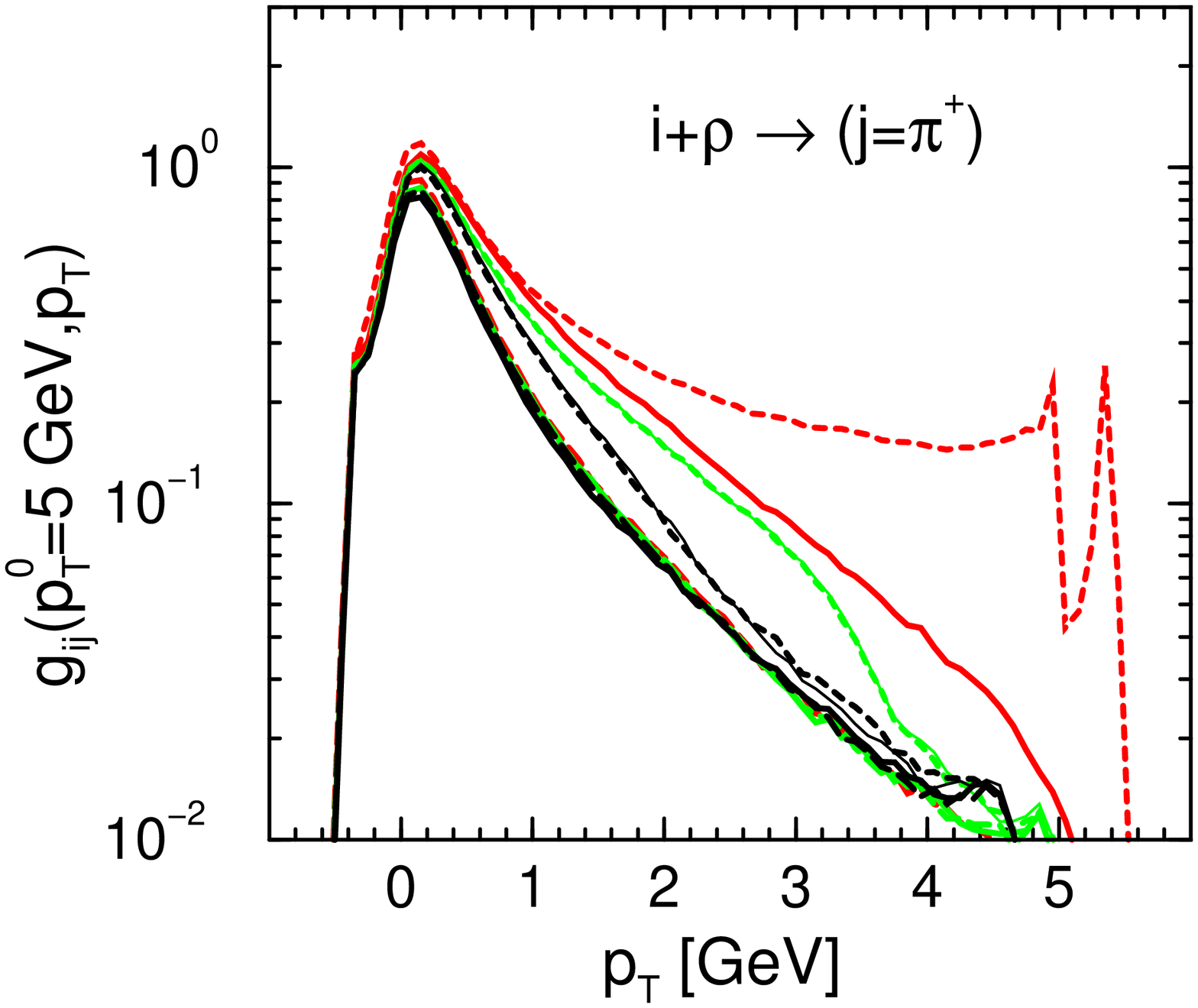}
\hspace*{\fill}
\includegraphics[angle=-90,width=6cm]{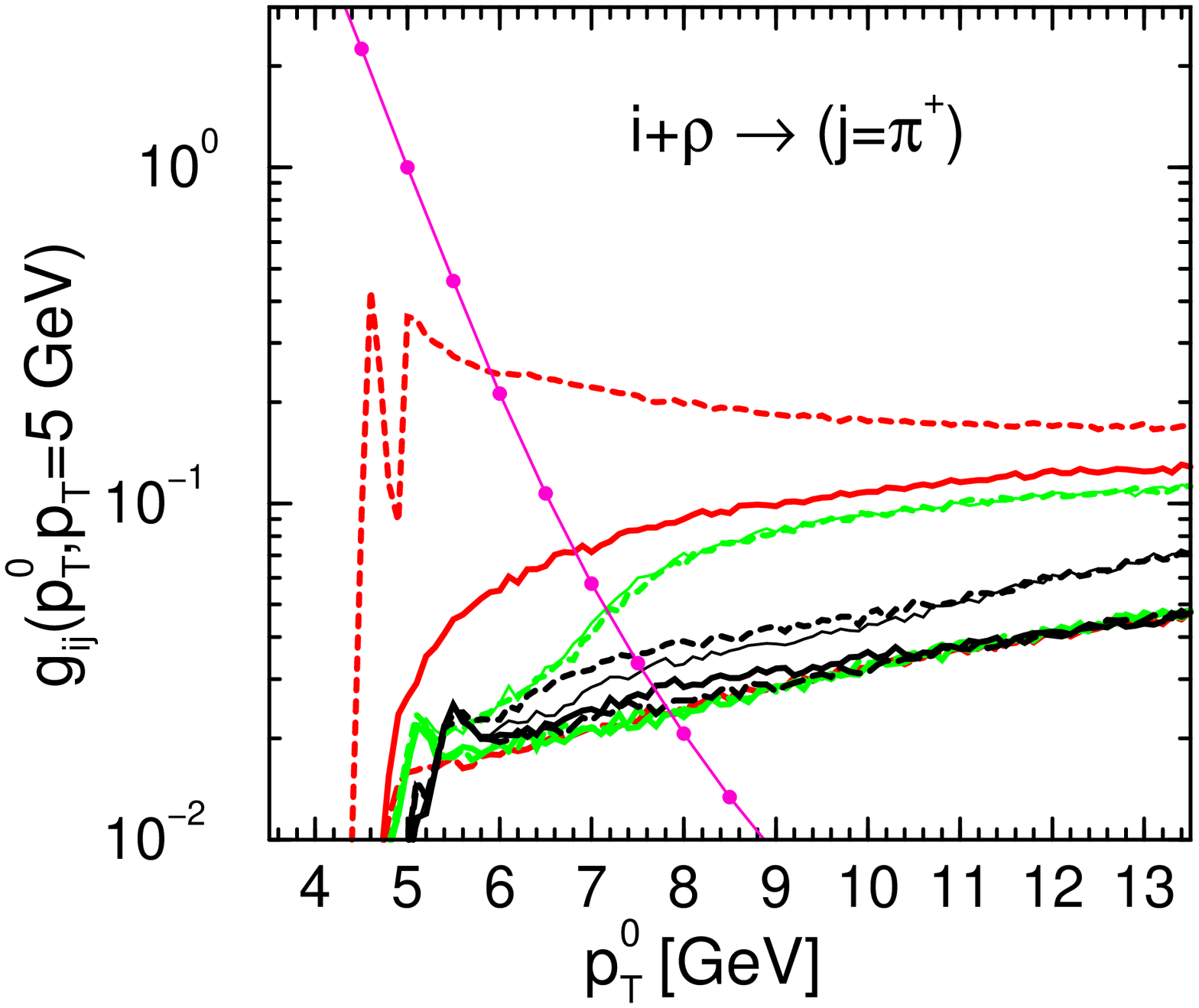}
\hspace*{\fill}

\caption{The folding matrices $g$ in (1) according to
      (in)elastic scattering on a $\rho$ for a final $\pi^+$ meson
      as a function of the resulting
      transverse momentum $\pT$ (left). Here a fixed value of the initial
      transverse momentum $\pT^0=5\GeV/c$ has been assumed. The r.h.s.
      shows the folding matrices $g$  as a function
      of the initial transverse momentum $\pT^0$ for a given
      final  transverse momentum $\pT=5\GeV/c$ (right).
      Pions, kaons and protons/neutrons
      are separated by line colour, while their charge state
      [negative (long dashed), neutral (solid), positive (dashed)] is
      indicated by the line style. Neutral anti--particles are
      displayed by a thin line.}
\label{fmatr}
  \end{center}
\end{figure}

On the other hand, it is not clear presently to which
extent this suppression might be due to ordinary hadronic final
state interactions (FSI) \cite{Kai}, too. We have
recently proposed a scenario that (in-)elastic collisions of
(pre-)hadronic high momentum states with the bulk of hadrons
in the late fireball can also contribute significantly
 to the attenuation of the high transverse
momentum hadrons at RHIC (for an illustration  see Fig.~\ref{fig:scheme}).
The first idea is that most of the (pre--)hadrons stemming
from a jet might still materialize in the dense system
for transverse momenta up to 10 GeV/c. Indeed, the time for color
neutralization of the leading particle due to gluon emission
can be very small \cite{Kopel4}. The late hadronic
final state interactions with the bulk of comovers then
have a clear and nonvanishing effect in suppressing the spectrum \cite{Kai}.
This comes about because (in)elastic reactions
of the (pre-)hadrons with hadrons of the bulk system
at the relevant energy scale $\sqrt{s}$ of a few GeV
are strong and cannot be described by pQCD methods.
For a single collision the momentum degradation
can be calculated via the folding equation \cite{Kai}
\begin{equation}
  f_j(p_\perp) = \sum_i\int d\pT^0\ f^0_i(\pT^0)\ g^X_{ij}(\pT^0,\pT)\ .
  \label{eq:model0}
\end{equation}
In (1) $g^X_{ij}(\pT^0,\pT)$  indicates the probability that
from a given particle
$i$  with transverse momentum $\pT^0$ one gets a particle $j$ with
transverse momentum $\pT$, if the collision is taken with a
target $X$ at rest.
The folding matrix can be modeled via the FRITIOF scheme;
typical examples are depicted in Fig.~\ref{fmatr}.
Such (in-)elastic collisions are very efficient for energy
degradation since many hadrons with lower energies are produced.
On the average 1 to 2 such interactions can account already
quantitatively for the attenuation of high $\pT$ hadrons at RHIC \cite{Kai}.

Hence, the  hadronic attenuation has therefore to be
addressed in more detail before conclusions on possible
QCD effects in a deconfined QGP phase on the materializing jets
can be  drawn. For a  quantitative
analysis (cf. \cite{CGG}) we employ the HSD transport approach \cite{Ehehalt}.

\begin{figure}
\begin{center}
    \includegraphics[width=6cm]{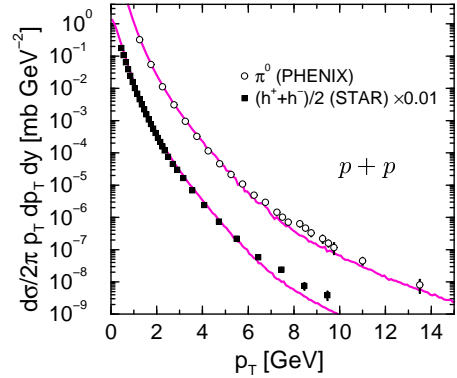}

    \caption{
      The invariant cross sections for the production of neutral pions
      and charged hadrons in p+p collisions (\SqrtS{200}) at
      midrapidity as a function of transverse momentum $\pT$.
      The \Pythia{} (v6.2) calculations (solid lines) are
      compared to the experimental data from RHIC.
      }
    \label{fig1}
  \end{center}
\end{figure}

\begin{figure}
  \begin{center}
\includegraphics[width=6cm]{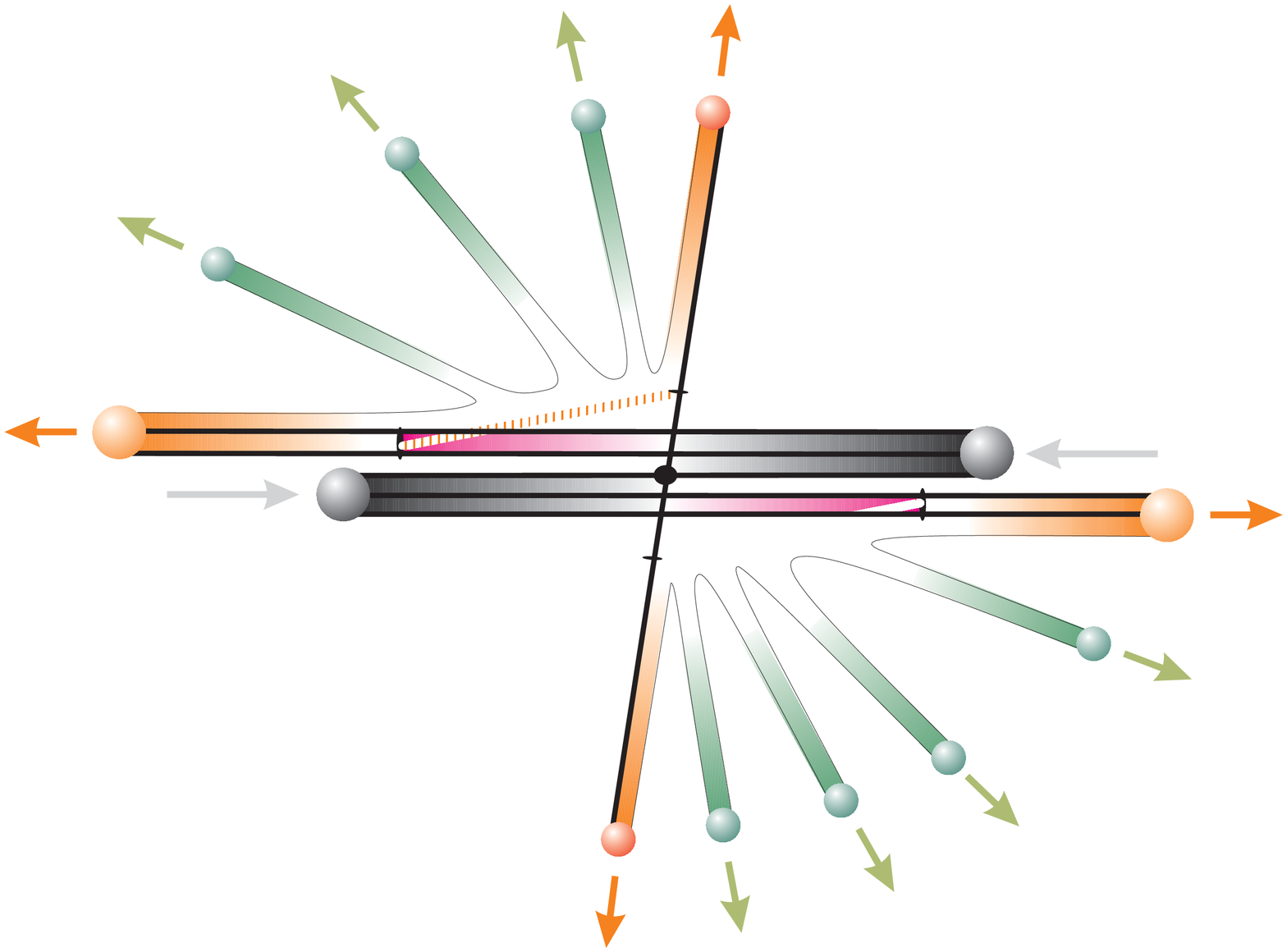}
    \includegraphics[width=6cm]{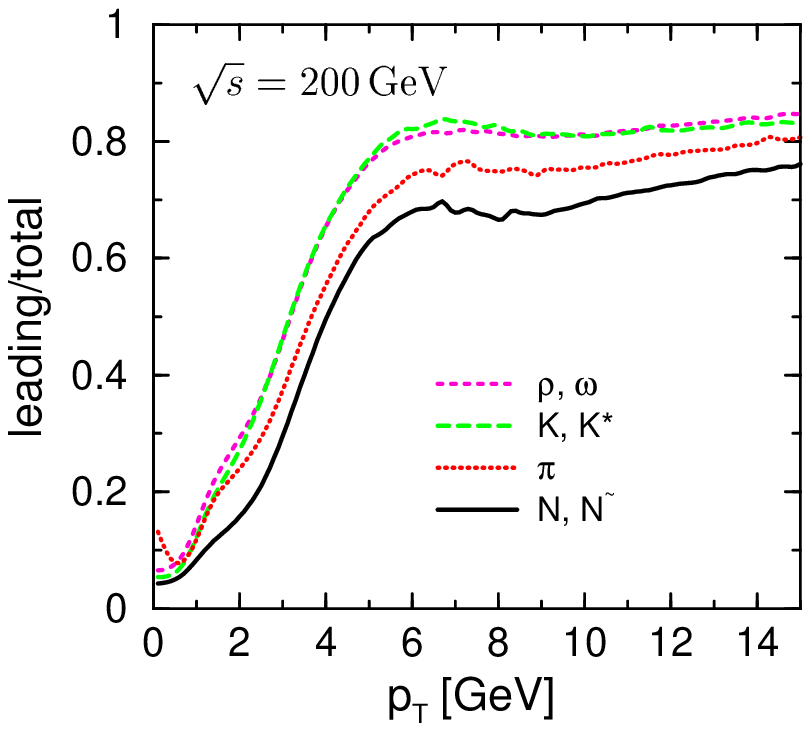}

    \caption{
Left panel: Sketch of a nucleon--nucleon collision. Particles containing a
string end (di--)quarks are denoted by ``leading'' (pre--)hadrons and may interact
immediately (with a reduced cross section), while the ``secondary''
hadrons are delayed by a formation time $\tau_f$.$ \quad $
Right panel:
      The ratio of 'leading particles' to 'all produced particles'
      in N+N collisions
      at midrapidity for different particle classes as a function
      of the transverse momentum $p_T$ within the \Pythia{} description.
      }
\label{sketch}
    \label{fig3}
  \end{center}
\end{figure}

The initial conditions for the production and also
subsequent propagation of hadrons with moderate to high transverse
momentum ($>1.5\GeVc$) are incorporated by a superposition of
$p+p$ collisions described via PYTHIA \cite{PYTHIA}, which serve as
the basic input and have been adjusted to the experimental data for $pp$ ractions
(cf. Fig.~\ref{fig1}).
Before coming to the actual results we  briefly explain the
concept of 'leading' and 'secondary' hadrons in the transport
approach (see Fig.~\ref{sketch} for an illustration). In a
high energy nucleon-nucleon collision two (or more) color-neutral
strings are assumed to be formed. The string ends are defined by the
space-time coordinates of the constituents.
These constituent quarks (or diquarks or
antiquarks) are denoted as 'leading' quarks that constitute the
'leading' pre-hadrons.
The time that is needed for the fragmentation of the strings and
for the hadronization of the fragments is denoted as {\it formation
time} $\tau_f \approx 0.8$ fm. Due to time dilatation the formation
time $t_f$ in any reference frame is then proportional to the
Lorentz $\gamma$-factor.
We assume  that hadrons, whose constituent quarks and antiquarks
are created from the vacuum in the string fragmentation, do {\it
not interact} with the surrounding nuclear medium within their
formation time $t_f$.
On the other hand, for the leading pre-hadrons, i.e.~those involving
quarks (antiquarks) from the struck nucleons, we adopt a reduced
effective cross section $\sigma_{lead}$ during the formation time
$t_f$ and the full hadronic cross section later on.
As it turns out \cite{CGG}, hadrons with transverse
momenta larger than $\sim 6\GeVc$ predominantly stem from the
string ends and therefore can, in principle,
interact directly with the reduced cross section (cf. Fig.~\ref{fig3}).
In addition, when turning to Au+Au collisions,
the formation of secondary hadrons is
not only controlled by the formation time $\tau_f$, but also by
the energy density in the local rest frame, i.e., hadrons are not
allowed to be formed if the energy density is above $1\UNIT{GeV/fm^3}$.
The interaction of the leading and energetic (pre-)hadrons
with the soft hadronic and bulk matter is thus explicitly modeled to occur
only for local energy densities below that cut.

Especially the interactions of leading (pre--)hadrons are important to
understand the attenuation of hadrons with high (longitudinal)
momentum in ordinary cold nuclear matter. The studies in Ref.
\cite{Falter} (see also \cite{Kopel4}) have shown that the
dominant final state
interactions of the hadrons with maximum momentum -- as
measured by the HERMES Collaboration \cite{HERMES} -- are
compatible with the concepts described above.

As a last prerequisite we note that a phenomenological
broadening of the average transverse momentum squared $\kTaveSq$ of the
partons in the nuclear medium prior to the 'hard' scattering
vertex is incorporated via an increasing width
\begin{equation}
\label{kt}
\kTaveSq = \kTaveSq_{pp} (1+\alpha N_{prev})
\end{equation}
in the string fragmentation function for a given number of
previous collisions $N_{prev}$. This modelling
mimics the so called `Cronin effect' observed in
$p+A$ collisions.
The parameter $\alpha \approx 0.25-0.4$ is fixed \cite{CGG} in comparison to
the experimental data for d+Au collisions at $\sqrt{s}$ = 200 GeV \cite{E1,E2}.

\section{Results of Transport calculations for Au+Au collisions}

\begin{figure}[htb!]
  \begin{center}
    \includegraphics[width=7cm]{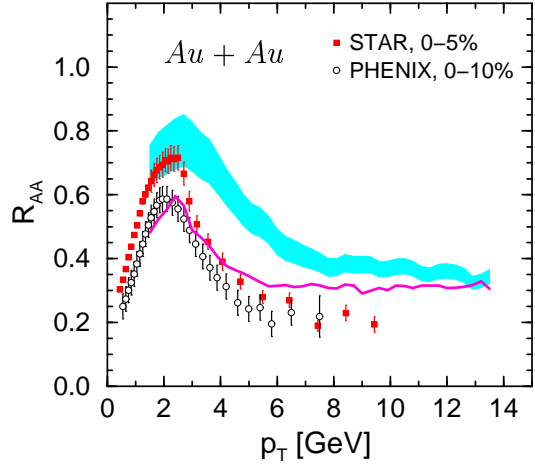}

    \caption{
      The suppression factor $R_{\rm AA}$ (3) of charged hadrons
      at $0\cdots5\proz{}$ central Au+Au collisions (\SqrtS{200})
      at midrapidity;
      experimental data are from
      Refs.~\cite{PHENIX1,STAR1}.
      The hatched band denotes our calculations including the 'Cronin'
      effect, while the solid line results from transport calculations
      without employing any initial state Cronin enhancement.
      }
    \label{fig7}
  \end{center}
\end{figure}

Fig.~\ref{fig7} shows our results for most central (5\% centrality)
Au+Au collisions at \SqrtS{200} for the nuclear modification
factor
\begin{equation}
  \label{ratioAA}
  R_{\rm AA}(\pT) = \frac{1/N_{\rm AA}^{\rm event}\ d^2N_{\rm AA}/dy d\pT}
  {\left<N_{\rm coll}\right>/\sigma_{pp}^{\rm inelas}\ d^2
    \sigma_{pp}/dy d\pT}\ .
\end{equation}
We emphasize, that the Cronin enhancement is visible
at all momenta, but does not show up to be responsible for the peak
structure in the enhancement around $2\GeVc$.
As demonstrated in Ref. \cite{CGG} the direct pions show a reduced attenuation, the kaon
reduction is slightly larger for lower $\pT$,
 while the vector meson absorption is much stronger.
Hadron formation time effects do play a substantial role
in the few $\GeVc$ region since heavier hadrons are formed earlier
than light pions in the cms frame at fixed transverse momentum due
to the lower Lorentz boost. The interactions of formed hadrons
after a formation time $t_f$  are not able to explain the attenuation observed
experimentally for transverse momenta $\pT\geq6\GeVc$. However,
the shape of the ratio $R_{\rm AA}$ in transverse momentum $\pT$
reflects the presence of final state interactions of formed
hadrons in the $1\dots5\GeVc$ range \cite{CGG}.
Such a behaviour for moderate transverse momentum
has also been demonstrated in an
exploratory UrQMD simulation in Ref. \cite{Bass}.

As pointed out before, the suppression seen in the calculation
for larger transverse momentum hadrons is due to the interactions
of the leading (pre-)hadrons with target/projectile nucleons and
the bulk of low momentum
hadrons. For the most central collisions, still, the experimentally
observed suppression can not fully be described by the (pre-)hadronic
attenuation of the leading particles. The ratio $R_{\rm AA}$ (3) decreases to a value
of about $0.35-04$ for central collisions, whereas the data range between
$0.2 -0.25$.
Our calculations, though,  turn out to be in a better agreement with the data
for more peripheral reactions than for the most central
Au+Au collisions \cite{CGG}.

\begin{figure}[htb!]
  \begin{center}
    \includegraphics[width=7cm]{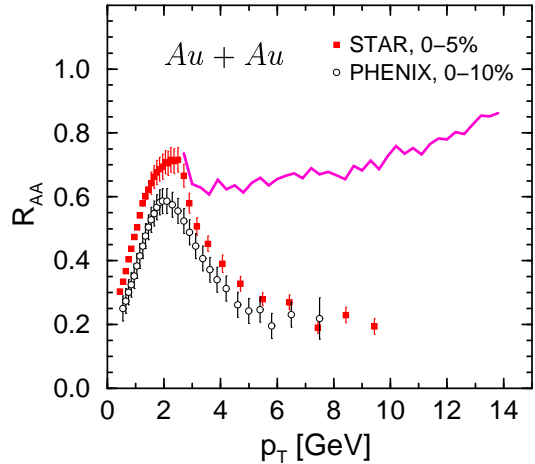}

    \caption{
      Same as fig.~\ref{fig7}, but with a leading cross section
      according to eq.~(\ref{sigma}) for the perturbative high $\pT$
      particles.
      }
    \label{fig10}
  \end{center}
\end{figure}

As an interlude, we turn briefly to an alternative model for the leading pre-hadron cross
section since the notion of a fractional constant cross section might be
questionable \cite{Kopel4} and alternative assumptions should be tested.
To this aim we have adopted a time--dependent cross section for
leading pre-hadrons of the kind
\begin{equation}
  \label{sigma}
  \sigma_{\rm lead}(\sqrt{s},\tau) = \frac{\tau-\tau_0}{\tau_f} \sigma_{\rm had}
  (\sqrt{s})
\end{equation}
for $\tau-\tau_0 \leq \tau_f$, where $\tau_0$ denotes the actual production
time and $\tau_f$ its formation time in the calculational frame. The full
hadronic cross section is adopted for $\tau \geq \tau_0$. The numerical
results of this assumption are shown for the ratio $R_{\rm AA}$ (3) in
Fig.~\ref{fig10} in case of 5\proz{} central Au+Au collisions.
Within this scenario the attenuation is less than 40\% up to
$\pT\sim10\GeV/c$ and thus provides a noticable but not  dominant
contribution. Since the trend of high $\pT$ hadron suppression is better described
via the standard picture of leading (pre-) hadron cross sections,
as also employed for  the attenuation of maximum momentum
particles in nuclei at HERMES \cite{Falter}, we now return to this default
description.

\begin{figure}[htb!]
  \begin{center}
    \includegraphics[width=7cm]{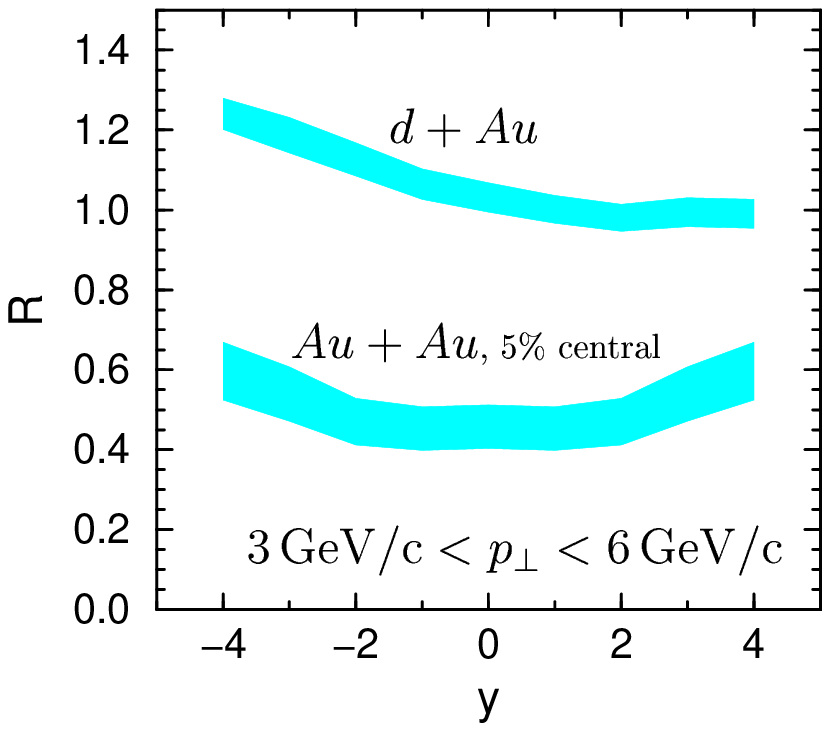}

    \caption{
      The suppression factor $R$ in $d+Au$ and $Au+Au$ collisions averaged
over the $\pT$ range $3\cdots6\GeV/c$ as a function of rapidity
$y$. The hatched bands indicate the uncertainty due to the initial
state Cronin effect.
      }
    \label{figyDep}
  \end{center}
\end{figure}

In Fig.~\ref{figyDep} we show the rapidity dependence of the
suppression factor $R$ averaged over the $\pT$ range $3\cdots6\GeV/c$ in
d+Au and Au+Au collisions.
While our results for Au+Au show a slight increase at higher
rapidities, one observes a noticable asymmetry with respect to midrapidity
in the d+Au case. This
is due to the fact, that all hadronic particles in d+Au reactions
are shifted to the
rapidity of the Au nucleus as also observed for the dominant
soft particles \cite{CGG}.

Gating on particles with momenta $\pT\geq4\GeVc$ in the transport
calculation, only 1/3 of these final hadrons have suffered one or more
interactions during their propagation to the vacuum, whereas the other 2/3
escape without any interaction (in  case of a central collision as
depicted in Fig.~\ref{fig7}).
This observation implies, since more than 3/4 of the high $\pT$
hadrons are strongly absorbed, that the final high $p_T$ hadrons seen
experimentally essentially stem from pre--hadrons that originate from
a diffuse 'surface region' of the expanding fireball.
The latter pre--hadrons then evolve (or fragment) to the final hadrons
dominantly in the vacuum and are accompanied by secondary hadrons,
which - due to their large formation time - also hadronize in the
vacuum.

\begin{figure}[htb!]
  \begin{center}
    \includegraphics[angle=-90,width=7cm]{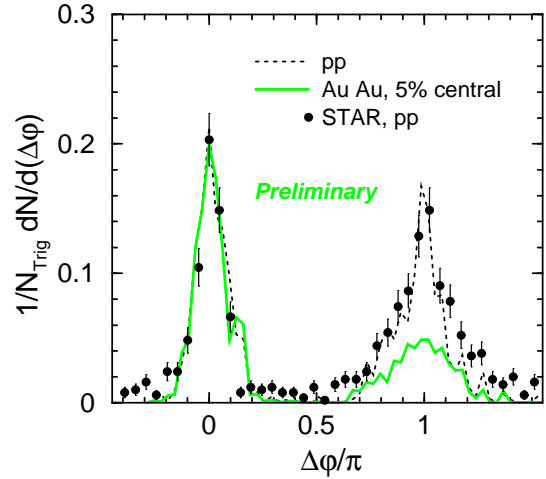}

    \caption{
      Near-side and far-side jet-like correlations from HSD for p+p and central
Au+Au collisions at midrapidity for $\pTN{Trig}=4\dots6\GeV/c$ and
$\pT=2\GeV/c\dots\pTN{Trig}$.
      }
    \label{angcorr}
  \end{center}
\end{figure}

In Fig.~\ref{angcorr} we show the angular correlation of high $\pT$
particles ($\pTN{Trig}=4\dots6\GeV/c$, $\pT=2\GeV\dots\pTN{Trig}$, $|y| <0.7$)
for 5\% central Au+Au collisions at $\sqrt{s}$ = 200 GeV (solid line)
as well as $pp$ reactions (dashed line) in comparison to the data
from STAR for $pp$ collisions \cite{StarAngCorr}.
Thus, when gating on high $\pT$ hadrons (in the vacuum) the
'near--side' correlations are close to the 'near--side'
correlations observed for jet fragmentation in the vacuum.
This is in agreement with the experimental observation \cite{StarAngCorr}.
Furthermore, for the far-side correlations we get
a $\sim$60\% reduction, but not a  complete disappearance
of the far-side jet as indicated by the experimental data \cite{StarAngCorr}.

\section{Summary, Conclusions and some Remarks}

Summarizing, we point out that (pre-) hadronic final state
interactions are able to approximately reproduce the high $\pT$
suppression effects observed in Au+Au collisions at RHIC.
This finding is important, since the same dynamics also describe
the hadron formation and attenuation in deep--inelastic lepton
scattering off nuclei at HERMES \cite{Falter} appreciably well.
In particular, it has been demonstrated, that the centrality dependence
of the modification factor $R_{\rm AA}$ 3 in Au+Au collisions at
$\sqrt{s}$ = 200 GeV is well described for peripheral and mid--central
collisions on the basis of leading pre-hadron interactions
\cite{CGG}. On the other hand, the attenuation in central Au+Au collisions is
 noticeably underestimated.
A similar situation also holds for an analysis
of the far-side correlations, which show a substantial, but
not fully complete suppression in central collisions as indicated
by the experimental data.
From these observations one should conclude that there
are some additional (and earlier) interactions of
partons in a possibly colored medium that have not been accounted for in
our present HSD transport studies. However, in any case, the final hadronic
stage is rather opaque for high transverse momentum and jet-like
(pre-)hadrons.

We note additionally, that the elliptic flow $v_2$ for high transverse
momentum particles is underestimated by at least a factor of 3 in our
transport calculations \cite{CGG}.
Moreover, also the experimental observation of having  more
protons than $\pi^+$ at moderate transverse momenta can not be
explained within the present approach. This, however, is true also
for the various parton jet--quenching models.

We close in pointing out that further experimental studies on the
suppression of high momentum hadrons from d+Au and Au+Au
collisions down to $\sqrt{s}$ = 20 GeV will be necessary to
clearly separate initial state Cronin effects from final state attenuation
and to disentangle the role of partons in a colored partonic medium
from those of  interacting pre-hadrons in a color-neutral hot and dense
fireball.

\end{document}